# Exposure Assessment for Wearable Patch Antenna Arrays at Millimeter Waves

Silvia Gallucci[1,2], Marta Bonato[1], Martina Benini[1], Marta Parazzini[1], , Maxim Zhadobov[3],

[1]Istituto di Elettronica e di Ingegneria dell'Informazione e delle Telecomunicazioni (IEIIT), Consiglio Nazionale delle Ricerche (CNR), 20133, Milan, Italy
[2]Dipartimento di Elettronica, Informazione e Bioingegneria (DEIB), Politecnico di Milano, 20133, Milan, Italy
[3]IETR/CNRS, University of Rennes, France

Corresponding author: Silvia Gallucci (e-mail: silvia.gallucci@ ieiit.cnr.it).

## Abstract

Since the spread of the wearable systems and the implementation of the forthcoming 5G in many devices, the question about the assessment of the exposure in wearable typical usage to millimeter waves is crucial and timely. For such frequencies, the power absorption becomes strongly superficial and involves only the most superficial tissue of the human body, i.e., the skin. In literature there are some models able to describe the layered structure of the skin but, until now, there is no literature consensus on the skin model to employ in computational exposure assessment studies. For these reasons, the present work aimed to simulate four different models of the most superficial tissues with different degree of detail exposed to two wearable patch antennas at different frequencies i.e., 28 GHz and 39 GHz. This allows to investigate the impact that the choice of a layered model rather than the homogeneous one has on the exposure. Simulations were performed through the FDTD method, implemented in the Sim4life platform and the exposure was assessed with the absorbed power density averaged over 1 cm$^2$ and 4 cm$^2$ ($S_{ab}$). The data showed that the homogeneous model underestimates the peak value of $S_{ab}$ obtained for multi-layer models in the stratum corneum (by 14% to 21% depending on the number of layers of the model and the frequency). This finding was confirmed by an analytical approach with two impinging plane wave TEM-polarized with normal incidence at 28 GHz and 39 GHz respectively. Conversely, there are no substantial differences in the exposure levels between the layered models.

*Keywords* – Wearable devices; computational dosimetry; skin models; millimeter waves.

## 1. Introduction

The use of wearable technologies is increasingly growing. They are very attractive for various applications, spreading from healthcare to smart home [1], [2]. The wearable technology is based on the concept of the Body-Area Network (BAN) consisting in a network of sensors/actuators/antennas able to exchange with each other and with an external gateway information about the user's health condition, position, environment and so on [3]. This involves communication between sensors, central BAN unit, and external node, e.g. smartphone, at two levels of communication: intra-BAN and inter-BAN [4]. The communication protocol of the wireless BAN (WBAN) is defined in the IEEE 802.15.6 standard [5] in which several frequency bands are mentioned. The operating frequencies include the 2.4 GHz Industrial-Scientific-Medical (ISM) band, which became standard for such type of systems due to the spread of the Bluetooth, BLE and Zigbee protocols [6].

Due to the way of use of the wearable devices that entails to pose them at very short distance from the human body, the assessment of the human exposure to the electromagnetic field (EMF) emitted by these devices is needed. Indeed, there is increasing public concern regarding safety issues for new/emerging wireless systems [7].

By merging the new 5G frequency bands, particularly the millimeter waves (mmWaves), and the wearable devices, the exposure assessment becomes even more necessary. In this regard, in literature there are some studies based on the design of novel mmWave wearable antenna in which the question about the human exposure is addressed (see as examples [8], [9]) but these studies do not appear exhaustive both in terms of the quantities used to assess the exposure and the used human models.

Since in the upper part of the microwave spectrum the absorbed power is confined in the superficial, to characterize numerically antenna/body interactions the use of appropriate tissue model is crucial as it directly impacts the reliability and accuracy of results. The anatomical human models typically used in dosimetric studies (e.g., [10]) do not represent the skin structure in sufficient details for mmWaves dosimetry. Indeed, in this type of models, the skin is typically modelled as a homogeneous tissue, disregarding its heterogeneous structure [11]. For this reason, stratified multi-layered cutaneous models were introduced in literature [7], [12].

Alekseev et al. [13] derived the dielectric properties of two skin layers at various locations on the body: the stratum corneum (SC), the external layer of skin, and the viable epidermis and dermis, the inner ones, in the frequency range 37-78 GHz. In another study [14], the same group compared three stratified skin models when they are exposed to a plane wave in the frequency range 30-300 GHz: the first one made of dermis, the second one made of SC and viable epidermis and dermis, and the third one based on three layers (SC, viable epidermis and dermis, and fat). In this study, the authors also investigated the impact of the skin thickness depending on the on-body location (forearm and palm). The results demonstrated that there are no differences between power density and specific absorption rate (SAR) in the homogeneous model compared to a layered model of thin (0.015 mm) SC. Sasaki et al. [15] used a skin model where viable epidermis and dermis were modelled as two separate layers and followed by a subcutaneous adipose tissue and a muscle layer. By means of Monte Carlo method they showed that, for a plane wave with frequencies ranging from 10 GHz to 1 THz, thickness variation affects the power absorption. Sacco et al. [16] considered the age-dependent variations of the skin permittivity and thickness. The models consisted of four layers: SC, viable epidermis and dermis, fat, and muscle. The results showed that the skin thickness variations affect the exposure for the lowest frequency (i.e., 26 GHz) and in particular for people < 25 years old; in general, considering both the analysed variations, the power transmission coefficient increases with age for both the considered frequencies. Christ et al. [17] focused on the temperature increase induced in a five-layered model: SC, viable epidermis, dermis, fat, and muscle. The results showed that the homogeneous model underestimates more than a factor of three the induced temperature increase with respect to the layered one, when plane wave frequencies (6-100 GHz) and thicknesses of the SC (10-700 µm) are varied. Later, Christ et al. [18] demonstrated that the homogeneous model with properties of dermis underestimates the transmission coefficient at the air/skin interface (T($\theta$), where $\theta$ is the incidence angle of the plane wave) compared to two multi-layered models made of SC, viable epidermis and dermis, fat, and muscle when the model is impinged by a plane wave in the frequency range of 6-300 GHz. These two layered models differ with each other for the thickness of the outer layer, i.e., the SC. Finally, Ziskin et al. [12] calculated the reflection coefficient, the power deposition, and the temperature increase in two different skin models: three- and four-layered ones. The models were tested with an incident plane wave at frequencies in the 37-78 GHz range and the results obtained by means of both the analytical and the computational approaches revealed that the power absorption is strongly localized on the most superficial layers, as expected. From the point of view of the type of the used model, they affirmed that the relevance in studies at such frequencies of the multi-layered models also including the inner tissues (i.e., fat and muscle) is linked to the thermal analysis since the heat propagates deeper compared to the EMF.

In light of the abovementioned literature studies, it is clear that there is no literature consensus on the skin modelling approach to employ in computational exposure assessment studies. Moreover, even the international organizations, responsible for the regulations aimed to the radiation protection, do not refer to a standard model for the skin. Indeed, the ICNIRP guidelines [19] refer to the absorbed power density as a dosimetric quantity at mmWaves without defining neither which is the appropriate model to reproduce the skin, nor which is the skin layer to considered for the comparison with the exposure limits. On the other hand, IEEE Std. C95.1 [20] suggests using the epithelial power density as the dosimetric quantity between 6 GHz and 300 GHz, referring with "epithelial" to the SC. In their recent review paper, Hirata et al. [7] reaffirmed the importance of using appropriate human models because the improvement of their degree of details can make the results obtained by means of the computational approach even more reliable. This is valid especially at mmWave frequencies, where a realistic representation of the skin structure could strongly impact on the exposure assessment.

For all these reasons, further investigations are needed with the aim to deepen the variation on the exposure results owing to the employed approach for the cutaneous tissue modelling. To the best of our knowledge, the literature is lacking in studies of exposure assessment due to mmWave frequencies and, particularly, to wearable antennas tuned to such frequencies, that evaluate skin tissue models with different stratifications.

This paper takes place in this context, aiming to simulate different planar geometrical layered models of the most superficial tissues with different degree of detail in the description of the cutaneous tissue exposed to two wearable patch antennas array at different frequencies both belonging to the 5G bands i.e., 28 GHz and 39 GHz. This allowed to investigate the impact that the choice of a layered model rather than the homogeneous one has on the exposure assessment. More specifically, four multi-layered models with increasing complexity were simulated: from a homogeneous model with dermis properties to the four-layered model composed of the SC, dermis, fat, and muscle. For each model, the exposure assessment was performed both by a computational approach making use of the FDTD method and an analytical approach with the estimation of the absorbed power density when the skin is hit by a normally incident plane wave at 28 GHz and 39 GHz.

## 2. Materials & Methods

This section is organized as follows. First, the tissue models are introduced in terms of their geometrical and electromagnetic properties. Then antenna design is presented, and the numerical method and analytical approach are described.

### 2.1 Anatomical Models

We considered four superficial tissue models of increasing complexity, from a homogeneous one to a stratified four-layered model. More in detail, the simulated models are: (i) homogenous single layer with dermis properties, (ii) two-layered, made of SC and dermis, (iii) three-layered made of SC, dermis, and fat, (iv) four-layered, modelled as SC, dermis, fat, and muscle [16]. The dielectric properties of each layer (Table I) were chosen according to the data found in literature at 30 GHz and 40 GHz [12], [17] and here assigned to 28 GHz and 39 GHz, respectively. The range of thickness of each layer were taken from the literature [12], [17]. With more details, the range of thicknesses of the SC were chosen in the range of dry "thin skin" because most of the body regions belong to it except for the palms and the soles of the feet. Table II report the thicknesses used here and obtained as belonging to the realistic range for the layers of fat, muscle, and viable epidermis and dermis; whereas the thickness of the stratum corneum was chosen within the ranges of thickness variation found in literature in order to optimize the models to have the same power transmission coefficient The maximum difference among the multi-layered models was of 4% for an impinging TEM-polarized plane wave with the incidence angle ($\theta$) varying from 0° up to 80°.

The overall dimension of the models is 150 x 150 mm, and the depth was chosen to be thick enough in order to neglect the possible contribution from the reflection at the deepest interface.

TABLE I
DIELECTRIC PROPERTIES AT 28 GHZ AND 39 GHZ

| Tissue | Relative Permittivity ($\varepsilon_r$) | | Conductivity ($\sigma$) [S/m] | | Density ($\varrho$) [kg/m³] |
|---|---|---|---|---|---|
| | 28 GHz | 39 GHz | 28 GHz | 39 GHz | |
| *Stratum Corneum (SC)* | 3.52 | 3.33 | 1.21 | 1.44 | 1500 |
| *Viable Epidermis & Dermis* | 16 | 12.1 | 27.5 | 32.6 | 1109 |
| *Fat* | 3.42 | 3.1 | 2.32 | 2.68 | 911 |
| *Muscle* | 21.5 | 15.6 | 39.9 | 47.1 | 1090 |

TABLE II
THICKNESSES [mm] OF THE LAYERS

| Tissue | Homogeneous | 2-Layered Model | 3-Layered Model | 4-Layered Model |
|---|---|---|---|---|
| *Stratum Corneum (SC)* | - | 0.02 | 0.02 | 0.02 |
| *Viable Epidermis & Dermis* | ∞ | ∞ | 0.96 | 0.96 |
| *Fat* | - | - | ∞ | 1.6 |
| *Muscle* | - | - | - | ∞ |

### 2.2 Antenna Models

The wearable antennas have to comply with constraints in terms of the compact size, lightness, and low profile [21]. To satisfy the aforementioned requirements, the two simulated antennas, inspired from Chahat et al. [22] and redesigned at 28 GHz and 39 GHz, are microstrip-fed four-patch antenna arrays. They consist of three different layers: ground plane, radiative element, and RT Duroid 5880 substrate ($\varepsilon_r = 2.2$, $\sigma = 5 \cdot 10^{-4}$ S/m). The overall antenna dimensions and inter-element distances are chosen to resonate at 28 GHz and 39 GHz, and they are detailed in Table III. Figure 1 shows the geometry of the antennas.

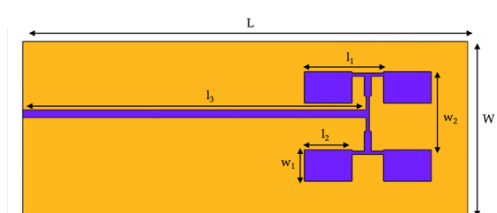

**Fig.1** The geometry of simulated wearable antennas.

TABLE III
DIMENSIONS [mm] OF THE ANTENNAS

| Values | Dimensions | |
|---|---|---|
| | 28 GHz | 39 GHz |
| L | 40 | 33.5 |
| W | 16.62 | 13.6 |
| $l_1$ | 9.6 | 6 |
| $l_2$ | 5 | 3.6 |
| $l_3$ | 30.25 | 25.88 |
| $w_1$ | 3.4 | 2.4 |
| $w_2$ | 9.6 | 6 |

### 2.3 Computational Approach

The exposure was assessed for the antenna located 2 mm from the model. Fig. 2 represents the positioning of the antenna with respect to the human model. The exposure scenario was the same for both antennas and the accepted power was set in both cases to 100 mW.

The EMF was computed using the Finite-Difference Time-Domain (FDTD) solver. Briefly, the FDTD method involves both a spatial and temporal discretization of the electric and magnetic fields over a period of time and a specific spatial domain limited with the boundary conditions. Typically, the minimum spatially sampling is at intervals of 10-20 per wavelength, and temporal sampling is sufficiently small to maintain stability of the algorithm [23]. All the simulations were performed in the software platform Sim4Life v.7 (ZMT Zurich Med Tech AG, Zurich Switzerland, www.zmt.swiss, accessed on 9 February 2023).

The computational domain was discretized with an automatic non-uniform grid for the antenna and the surrounding of the phantom with a sub-wavelength resolution of around 15 samples per wavelength. We set the mesh cell size varying from 0.06 mm to 0.33 mm depending on the dielectric properties of the model, in order to correctly discretize all the tissues to guarantee the compliance with the constraint of $\lambda/10$ imposed by the FDTD method for its stability. This resulted in a total of 8.537 and 162.310 MCells, respectively at 28 and 39 GHz. The computational domain was truncated by assuming 8 layers of perfectly matched layer (PML) material and 10 cells of free space were added around the computational domain at the domain boundaries.

All the simulations were performed on a workstation Z8 16-Core Processor @3.8 GHz, RAM 512 GB, with a mounted graphical card NVIDIA GeForce RTX5000. To speed up simulations, Sim4Life GPU accelerator aXware was used, and the maximum computational time was 30 minutes.

The Absorbed Power Density ($S_{ab}$, W/m$^2$) averaged over a surface of 1 cm$^2$ or 4 cm$^2$ of tissue was calculated as:

$$S_{ab} = \frac{1}{A} \int_S \mathrm{Re}\,[E \times H^*]\, dS \qquad (1)$$

where A is the surface area, dS is the surface element and Re [E x H*] is the real part of the Poynting vector.

According to the ICNIRP Guidelines [19], the surface area A on which such parameter has to be averaged is depending on the frequency. More in detail, the $S_{ab}$ averaged over 1 cm$^2$ is mandatory for frequencies greater than 30 GHz, since focal beam exposure can occur, whereas for lower frequencies the $S_{ab}$ averaged over 4 cm$^2$ is the adopted quantity. Indeed, in the ICNIRP guidelines the additional spatial average of 1 cm$^2$ is used to ensure that the operational adverse health effect thresholds are not exceeded even over smaller regions [19]. However, since the selected frequency of 28 GHz is close to 30 GHz, we have opted for the extraction of the $S_{ab}$ averaged over both the 1cm$^2$ and 4 cm$^2$ areas, for both frequencies.

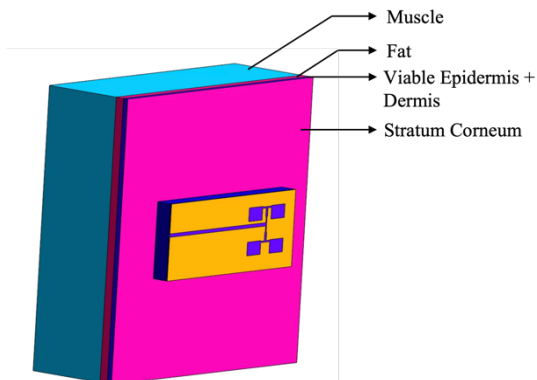

**Fig.2** Examples of simulated scenario with antenna positioned 2 mm way from the center of the skin model (for the sake of readability, the size of the phantom and antenna are not in scale).

### 2.4 Analytical Approach

In addition to the computational approach, the present study aimed also to expose the abovementioned superficial tissue models with a plane wave. More specifically, all the tissue models described before were exposed to TEM-polarized plane waves, the first one at 28 GHz, and the second one at 39 GHz, and the reflection coefficient was calculated at the interface air/most-external-layer following the generic formula (2) for a M-layered structure:

$$\Gamma_i = \frac{\varrho_i + \Gamma_{i+1} e^{-2jk_i l_i}}{1 + \varrho_i \Gamma_{i+1} e^{-2jk_i l_i}} \quad (2)$$

where i = M, M-1, …1 and it is initialized by $\Gamma_{M+1}=\varrho_{M+1}$. The so obtained $\Gamma$ was essential to calculate the absorbed power density ($S_{ab}$) by applying the formula (3) in the ICNIRP Guidelines [19]:

$$S_{ab}=(1-|\Gamma|^2) \cdot S_{inc} \quad (3)$$

where $S_{inc}$ is the incident power density, here imposed as 10 W/m² that is the reference level for general public exposure averaged over 30 minutes, the whole-body, and the frequency ranging from 2 to 300 GHz [19].

## 3. Results

The first section summarizes the results estimated through the computational approach with the averaged $S_{ab}$ peaks of each layer of the models, whereas the second section reports the results of the $S_{ab}$ calculated by means of the analytical approach through the calculation of the reflection coefficient ($\Gamma$) at the interface air/ most-external-layer.

### 3.1 Computational Approach

The computed peak values of the $S_{ab}$ averaged over 1 cm² and 4 cm² surface are presented in Fig. 3 for all the analyzed scenarios.

As expected, across all models and tissues the peak values of the $S_{ab}$ averaged over 1 cm² (top row) resulted always higher than the corresponding ones averaged over 4 cm² (bottom row). This trend is justified by the averaging operation itself. Indeed, in this specific situation where the peak is strongly localized, to average over a higher surface means to average over a larger number of lowest values, that has the effect to reduce the averaged value. For the sake of brevity, the results that will be commented henceforth are the $S_{ab}$ averaged over 1 cm² but the trend is the same for the peaks obtained averaging over a surface of 4 cm².

As reported in the upper left panel of Fig.3, the highest peak with the antenna at 28 GHz is found in the SC of the 4-layered model i.e., 13.8 W/m² and the maximum variation of peak $S_{ab}$ in the SC in multi-layered models is around 3%. The lowest value is obtained for 2-layered model whereas the exposure levels induced in 3- and 4-layered models are almost identical (< 2% deviation). Moreover, the exposure levels of dermis in multi-layered models are almost identical (< 1% deviation) whereas for the fat the results showed a maximum variation of 9% across the 3- and 4- layered models. Furthermore, by passing through the layers from the outer to the inner one, the peak $S_{ab}$ decreases confirming that lowest values are always in the most internal stratum, whatever it is.

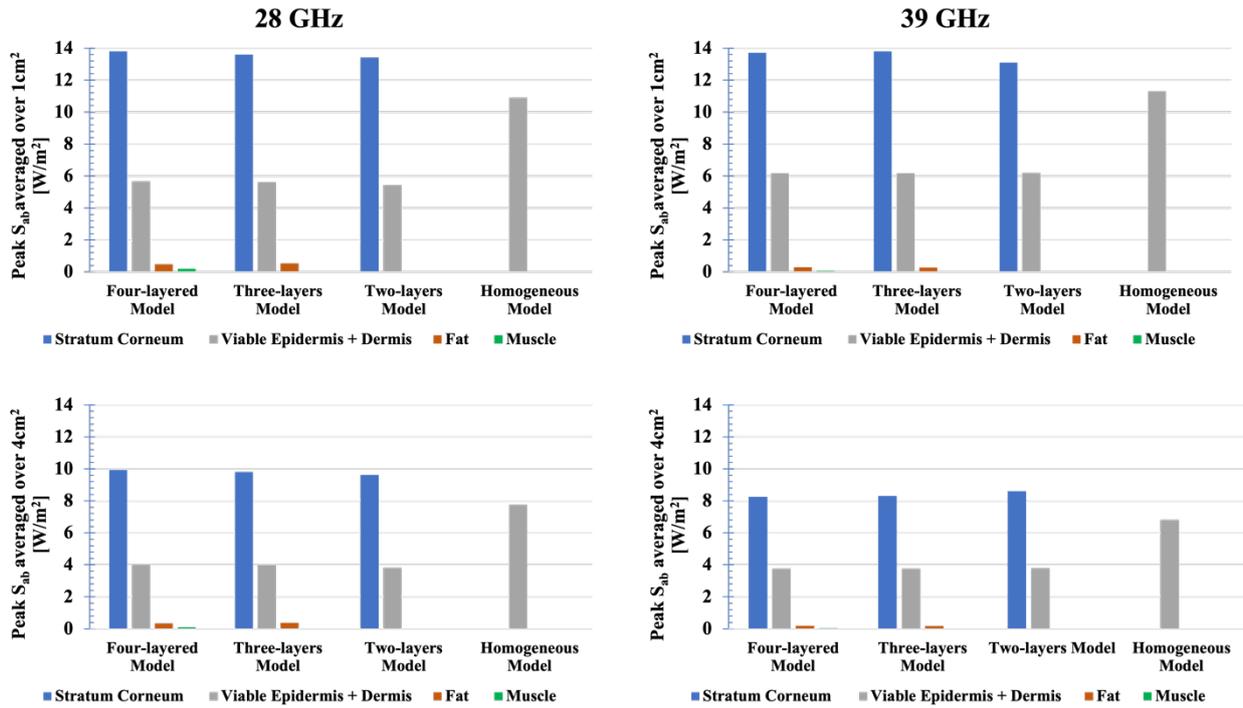

**Fig.3** Peak values of the absorbed power density ($S_{ab}$) for all the analyzed scenarios: on the left column the peaks of the $S_{ab}$ in the scenarios with the antenna tuned to 28 GHz, on the right column, the peaks obtained with the antenna tuned to 39 GHz; from the top to the bottom, the Sab averaged over 1 cm² and 4 cm², respectively.

As an example, the muscle, that is the fourth tissue in the 4-layered model, showed negligible exposure values reduced by around 98% with respect to maximum values at the SC.

The comparison between the responses of the models highlights that the homogeneous model strongly underestimates the peak value of $S_{ab}$ obtained for multi-layer models in the SC, i.e., in the most superficial layer. Indeed, the peak $S_{ab}$ obtained in the dermis of the homogeneous model is reduced by 18% to 21% depending on the number of layers, with respect to the highest exposure levels achieved in the SC of the multilayered models.

Nevertheless, from the comparison of the peaks $S_{ab}$ in the dermis across all the models, the difference between the peak in the homogeneous model (i.e., 10.9 W/m²) and the ones in the multi-layered models (i.e., 5.66 W/m², 5.63 W/m², and 5.44 W/m², from the four- to the two-layered models) is noteworthy. Certainly, this behavior is due to the presence in the multi-layered models of the SC that shields the EMF so reducing the power absorption in the inner strata.

On the panels of the right column, the peak values of the $S_{ab}$ over 1 cm² (top panel) and 4 cm² (bottom panel) when the antenna tuned to 39 GHz is simulated.

The general trend observed in the scenarios with the antenna tuned to 28 GHz is still valid in the second scenarios, in which the antenna tuned to 39 GHz is employed. First of all, the maximum peak is observed in the SC of the three-layered model i.e., 13.8 W/m².

Nevertheless, comparing this peak value with the ones of the 2- and 4-layered models there is no substantial differences, indeed the maximum deviation is around 5%. As for 28 GHz, the lowest peak $S_{ab}$ is observed in the 2-layered model whereas the comparison between the 3- and the 4-layered models did not reveal noteworthy variations (<1% deviation). Moreover, the peak $S_{ab}$ in dermis across the 2-, 3- and 4-layered models are almost the same (<1% deviation), whereas in the fat the variation between the 3- and the 4-layered models revealed almost 2% of deviation.

Certainly, the trend of the decrease of the exposure levels from the outer to the inner tissues is amplified in this case where a higher frequency is involved; indeed the peak $S_{ab}$ switches from 13.7 W/m² in the SC up to 0.07 W/m² in the muscle, showing a reduction of 99.5% of the exposure level.

Furthermore, the comparison between the multi-layered models and the homogeneous model showed the same behavior of the 28 GHz case: the homogeneous model (11.3 W/m²) tends to underestimate the $S_{ab}$ with respect to the peaks estimated in the SC of the multi-layered model; more specifically, the deviation varies from 14% to the maximum of around 18%. Conversely, the results obtained in the dermis of the multi-layered models (i.e., 6.17 W/m², 6.17 W/m², and 6.19 W/m² in the 2-, 3-, and 4-layered model respectively) are almost the same, showing a maximum variation of 0.3%.

### 3.2 Analytical Approach

In this section the results about the peak of absorbed power density ($S_{ab}$) calculated when an incident plane wave with normal incidence impacts on the most superficial tissues models are shown. The values are obtained for an incident power density ($S_{inc}$) imposed to the impinging wave equals to 10 W/m² according to the ICNIRP reference level for the general public in the frequency range 2-300 GHz [19].

Fig. 4 summarized the peak values of $S_{ab}$ calculated at the interface air/most-external-layer for both the frequencies and in all the models. More specifically, the here reported results are related to $S_{ab}$ absorbed by the SC for the 2-, 3-and 4-layered models and by the dermis for the homogeneous one.

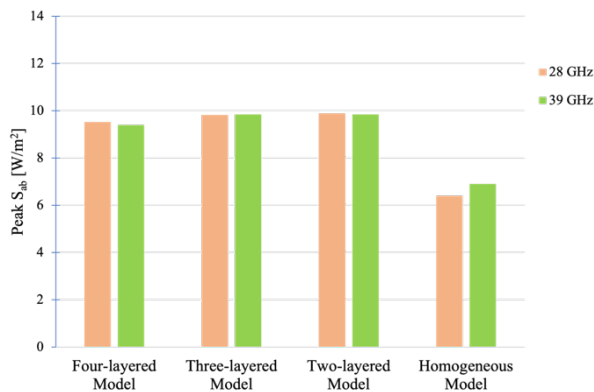

**Fig. 4** Peak values of the absorbed power density ($S_{ab}$) at the interface air/most-external layer of each layered model with the incident plane wave TEM-polarized with perpendicular incidence.

These results are in line with the previous ones obtained with the computational approach; indeed, for both frequencies the data shown that the homogeneous model still underestimates the $S_{ab}$ at the air/skin interface by 35% and 30%, for 28 and 39 GHz, respectively, when compared with the multi-layered models. Moreover, as before, the exposure levels in multi-layer models resulted almost identical (< 4% deviation across all the models for 28 GHz and < 5% deviation for the 39 GHz).

## 4. Discussion

Wearable wireless technologies are attractive for various communication and sensing applications, including personal healthcare, smart home, sport and so on. The healthcare has been the primary target application so far, however recently wearable communicating devices also demonstrated a potential for other usages such as military and entertainment [24]. The wearable devices may be a part of Wireless Body-Area Networks (WBAN) introduced in the IEEE 802.15.6 standard [5]. In WBAN, information deriving from the sensors is collected in a central unit and then transmitted to an external device (e.g., the smartphone) [4]. Recently, the wearable networks have also included 5G technology. Indeed, the use of the 5G protocol permits, for example, the possibility of the involvement of augmented, mixed, and virtual realities [24]. For this reason, 5G bands are involved in wearable communication, particularly in the mm-wave band (>24 GHz).

Since wearable devices are necessarily positioned on the human body, the question of the power absorbed by human tissues is crucial and timely, particular if mmWave wearable antenna are considered. Indeed, only few studies (see as examples [25]–[27]) aimed to assess the exposure generated specifically by wearable antennas in the 5G frequencies band using both simplified that detailed anatomical human models. In particular, only one very recent paper by Gallucci et al, [27] computationally assessed the human exposure due to the EMF emitted by wearable antennas, each one tuned to a 5G band (one tuned to f = 3.5 GHz and the second one to 26.5 GHz), positioned on the trunk of four realistic human models of the Virtual Population [10].

However, to numerically characterize antenna/body interactions, the use of appropriate tissue model is crucial as it directly impacts the accuracy of results. Indeed, particularly for mmWave frequencies up to 100 GHz, modelling the skin by a single layer of homogeneous dermis tissue with constant dielectric properties over its entire thickness, as it is done in the most popular anatomical models [10], could be an oversimplification to realistically represent the skin structure. As a consequence, stratified multi-layered models were introduced in literature [12], [13] These models are typically composed of the stratum corneum, dermis, fat, and muscle. However, in literature there is not yet a consensus about the approach to employ in studies of computational exposure assessment to model the cutaneous tissue.

This work is inserted in this context, investigating the exposure levels induced by two wearable patch antennas tuned in the mmWave bands, using models with different stratifications to investigate the impact that the choice of a multi-layered model rather than the homogeneous one has on the exposure assessment. Specifically, four planar models with increasing complexity were considered: from a homogeneous model with dermis properties to the four-layered model composed of the SC, dermis, fat, and muscle. The exposure was quantified by the assessment of the $S_{ab}$ averaged over both 1 cm² and 4 cm².

Analyzing the data of the peak value of the $S_{ab}$ when the antenna is tuned to 28 GHz, it is observed that the use of the homogeneous skin model led to an underestimation of the exposure level in the most external layer of the model with respect to the multi-layered models ranging from 18% to 21%, according to the 2-3- or 4-layered models. This trend is similar to the scenario with the antenna at 39 GHz showing an underestimation ranging from 14 % to 18%. In parallel, the analytical results confirmed the tendency found through the computational approach. Indeed, here the homogeneous model underestimates the exposure of 35.3% for the configuration with the antenna at 28 GHz, and 29.9% with the antenna at 39 GHz. Moreover, the grouping of these results by frequency shows the fact that the lower the frequency, the more noticeable the underestimation of the homogeneous model over the stratified models. This evidence is confirmed by the studies in literature, even though their number is limited; firstly, Bonato et al. [28] simulated the homogeneous, and the three- and four-layered models in three different exposure configurations to a 5G mobile phone antenna at 27 GHz (by varying the distance antenna-user), showing that the homogeneous model tended to underestimate the exposure in all the scenarios. Sasaki et al. [29] used the Monte Carlo simulation approach with varying the tissue thicknesses of the homogeneous and two-layered models. All the planar multi-layered models were hit by plane waves at frequencies from 0.1 to 1 THz and with 1 W/m² of incident power density. In their work, they demonstrated that the power transmittance increases when the skin is deeper modeled. Finally, Christ et al. [17] conducted a study in which incident plane waves at frequencies from 6 to 100 GHz impact on several stratified most superficial tissue models. Here, the reflection coefficient and the temperature increase were studied. This work highlighted the same trend of underestimation by the homogeneous dermis model by more than a factor of three, confirming the trend found in the present work and in the abovementioned studies.

Comparing the responses of the different multi-layered models, our data suggested that there are no substantial differences between the multi-layered models, particularly for the most external layer. In this regard, it was found that for the scenario with the lowest frequency the maximum variation of $S_{ab}$ is of 9% and it was between the four- and the three-layered models, precisely in the fat, whereas the greatest variation in the SC is of 3% between the two- and the four-layered models. This means that, with this frequency, choosing a stratified model rather than another always structured, the maximum expected impact on the exposure is 9%, particularly in the inner layer. For what concern the 39 GHz scenario, this maximum variation resides in the SC, and it is 5%, reducing the impact that the choice of a certain stratified model has on the exposure assessment, whereas for the inner strata the variation is almost 2%.

Finally, the comparison of the values reported in the left column with the peaks in the right column of Fig.3 showed that the peaks in the inner tissues (i.e., muscle) assessed in the scenario with the lowest frequency are higher than the peaks $S_{ab}$ observed in the case with 39 GHz. Indeed, the difference between the peaks with the antenna at 28 GHz and the ones with 39 GHz is more evident in the inner strata rather than the outer layers so much so that the variation in the SC is of 0.7%, whereas in the muscle it is 66.4%. This is in line with the decrease of the penetration depth corresponding to the increase of the frequency.

Overall, from the comparison with the IEEE International Guidelines [19], in any of the studied configurations the limit of 20 W/m² is exceeded, neither on the interface air/skin nor in the inner layers. However, the study was focused on the question about the best way to represent the most superficial tissue and, in light of the here presented findings, it is evident the difference between the exposure levels in the first layer of the multi-layered models (i.e., SC) and in the dermis of the homogeneous model, with an underestimation of almost 20% by the homogeneous model, used in most of the exposure assessment studies so far. This trend brings to the light the necessity to clearly define which is the layer where it is appropriate to estimate the power absorption because the correct exposure assessment derives from this definition.

Finally, the present study confirmed the results found in literature in which the homogeneous model underestimates the exposure levels and, moreover, expands these findings to more complex scenarios, no longer with an impinging plane wave, but with two real wearable antennas.

## 5. Conclusions

In conclusion, the present work aimed to assess the exposure due to two different mmWave wearable antennas using four most superficial tissues models of increasing complexity to investigate their effect on the exposure level.

The problem was addressed through both the computational and the analytical approach and both of them revealed the same considerations: comparing the $S_{ab}$ estimated at the most external layer of all the models, for both the frequencies, the peak in the homogeneous model is always lower than the ones of the layered models. This finding means that the use of the homogeneous skin model in exposure assessment studies with such high frequencies could underestimate the exposure, if compared with the high-detailed skin model.

## Acknowledgement

The authors wish to thank ZMT Zurich MedTech AG (www.zmt.swiss, accessed on 21 February 2023) for having provided the simulation software Sim4Life. The authors wish to thank the European Defence Agency (EDA) for the support to this work in the context of the project Nº. B 0987 IAP2 GP "Biological Effects of Radiofrequency Electromagnetic Fields (RFBIO)" funded by the Italian MoD.